\documentclass{IEEEtran}

\usepackage{cite}
\usepackage{amsmath,amssymb,amsfonts}
\usepackage{graphicx}
\usepackage{textcomp}
\usepackage{color,soul} 

\newcommand\tblu[1]{\textcolor[rgb]{0.00,0.00,0.00}{#1}} 

\def\BibTeX{{\rm B\kern-.05em{\sc i\kern-.025em b}\kern-.08em
    T\kern-.1667em\lower.7ex\hbox{E}\kern-.125emX}}
    
\begin{document}
\bstctlcite{IEEEexample:BSTcontrol}

\title{A multi-functional \tblu{reconfigurable} metasurface: Electromagnetic design accounting for fabrication aspects}

\author{
Alexandros Pitilakis,
Odysseas Tsilipakos, \IEEEmembership{Senior Member, IEEE},
Fu Liu, \IEEEmembership{Member, IEEE},\\
Kypros M. Kossifos, \IEEEmembership{Member, IEEE},
Anna C. Tasolamprou,
Do-Hoon Kwon, \IEEEmembership{Senior Member, IEEE},\\
Mohammad Sajjad Mirmoosa,
Dionysios Manessis,
Nikolaos V. Kantartzis, \IEEEmembership{Senior Member, IEEE},\\
Christos Liaskos, \IEEEmembership{Member, IEEE},
Marco A. Antoniades, \IEEEmembership{Senior Member, IEEE},
Julius Georgiou, \IEEEmembership{Senior Member, IEEE},
Costas M. Soukoulis,
Maria Kafesaki, and
Sergei A. Tretyakov, \IEEEmembership{Fellow, IEEE}

\thanks{Manuscript submitted March 2020; revised and accepted July 2020; published online 19 August 2020. This work was supported by the European Union Horizon 2020 Research and Innovation Programme-Future Emerging Topics (FETOPEN) under Grant 736876 (project VISORSURF). (\emph{Corresponding author: Alexandros Pitilakis}, email: alexpiti@auth.gr) }
\thanks{A. Pitilakis and N. V. Kantartzis are with the Department of Electrical and Computer Engineering, Aristotle University of Thessaloniki, 54124 Thessaloniki, Greece, and with the Institute of Electronic Structure and Laser, Foundation for Research and Technology Hellas, 71110, Heraklion, Greece.}
\thanks{O. Tsilipakos, A. C. Tasolamprou, C. M. Soukoulis, and M. Kafesaki are with the Institute of Electronic Structure and Laser, Foundation for Research and Technology Hellas, 71110, Heraklion, Crete, Greece. C. M. Soukoulis is also with the Ames Laboratory---U.S. DOE and Department of Physics and Astronomy, Iowa State University, Ames, Iowa 50011, USA.}
\thanks{F. Liu, M. S. Mirmoosa, and S. A. Tretyakov are with the Department of Electronics and Nanoengineering, Aalto University, P.O. Box 15500, FI-00076 Aalto, Finland.}
\thanks{K. M. Kossifos, M. A. Antoniades, and J. Georgiou are with the Department of Electrical and Computer Engineering, University of Cyprus, Cyprus. M. A. Antoniades is also with the Department of Electrical, Computer \& Biomedical Engineering, Ryerson University, ON, M5B 2K3, Canada}
\thanks{D.-H. Kwon is with the Department of Electrical and Computer Engineering, University of Massachusetts Amherst, Amherst, MA 01003, USA.}
\thanks{D. Manessis is with System Integration and Interconnection Technologies, Fraunhofer IZM, Berlin, Germany.}
\thanks{C. Liaskos is with the Institute of Computer Science, Foundation for Research and Technology Hellas, Heraklion, Greece, and with the University of Ioannina, Computer Science Engineering Department, Greece.}
\thanks{Digital Object Identifier (DOI): 10.1109/TAP.2020.3016479}
}

\markboth{IEEE TRANSACTIONS ON ANTENNAS AND PROPAGATION | DOI: 10.1109/TAP.2020.3016479 | \MakeLowercase{https://ieeexplore.ieee.org/document/9171580}}{Pitilakis \MakeLowercase{\textit{et al.}}}

\maketitle

\begin{abstract}
In this paper we present the theoretical considerations and the design evolution of a proof-of-concept \tblu{reconfigurable metasurface, primarily} used as a tunable microwave absorber, but also as a wavefront manipulation and polarization conversion device in reflection. 
We outline the design evolution and all considerations taken into account, from the selection of patch shape, unit cell size, and substrate, to the topology of the structure that realizes the desired tunability. 
The presented design conforms to fabrication restrictions and is co-designed to work with an integrated circuit chip for providing tunable complex loads to the metasurface, using a commercially available semiconductor process. 
The proposed structure can perform multiple tunable functionalities by appropriately biasing the integrated circuit: Perfect absorption for a wide range of incidence angles of both linear polarization states, accommodating a spectral range in the vicinity of $5$~GHz, \tblu{with potential also for wavefront control, exemplified via anomalous reflection and polarization conversion}.
The \tblu{end vision is for such a design to be scalable and deployable as a practical HyperSurface}, i.e., an intelligent multi-functional metasurface capable of concurrent reconfigurable functionalities: absorption, beam steering, polarization conversion, wavefront shaping, holography, and sensing.
\end{abstract}


\section{Introduction}
\label{sec:1:introduction}
\IEEEPARstart{R}{econfigurable} metasurfaces are electromagnetically ultra-thin sheets which are able to deploy tunable electromagnetic (EM) functions on demand. They consist of subwavelength elementary units, the meta-atoms, which can be engineered to enable interactions between the metasurface and the incoming wave \tblu{\cite{Soukoulis:2011,Glybovski:2016,Chen:2016IOP,Ma2019}}. This meta-atomic level manipulation has opened the path to the realization of a plethora of EM functions and applications involving wavefront shaping~\cite{Decker:2015,Tasolamprou:2015,Asadchy:2016PRB,Chalabi:2017,Tsilipakos:2018AOM,Wong:2018}, polarization control~\cite{Gansel:2009,Grady:2013,Yang:2014ACS}, dispersion engineering~\cite{Dastmalchi:2014,Tsilipakos:2018ACS}, perfect, broadband and asymmetric absorption~\cite{Aydin2011, Radi:2015PRApp,Liu:2017OpEx,Wang:2018,Perrakis:2019}, holography~\cite{Li:2017}, \tblu{imaging~\cite{Li2019}}, non-reciprocity \cite{Sounas:2013NatCom}, extreme energy accumulation~\cite{Mirmoosa:2019}, wireless power transfer~\cite{Liu:2019WPT}, harmonic generation~\cite{Li:2015}, etc. The key element in a reconfigurable metasurface is to have an inclusion (material or component) whose EM properties can be modified by an external stimulus. Following this rationale, various tunable electromagnetic functions have been demonstrated in metasurfaces that comprise stimulus-sensitive materials such as graphene~\cite{Wang:2017,Tasolamprou:2019}, liquid crystals~\cite{Shrekenhamer:2013}, and photoconductive semiconductors~\cite{Kafesaki:2012}. {In the communications community, the use of tunable metasurfaces, also referred to as reconfigurable intelligent surfaces (RIS), for control and optimization of the wireless propagation environment is actively discussed~\cite{Basar:2019b,Renzo:2019}}.

Most of the initial implementations referred to globally tunable metasurfaces, imposing limitations related to the lack of local control over the metasurface impedance. Clearly, for a multi-functional reconfigurable metasurface, an independently varying element is required in each unit cell to enable local control and  reconfigurability. A practical solution to this problem, highly suited to microwave realizations, is given by incorporating lumped electronic elements in the meta-atoms. In this way, the local impedance of the metasurface can be modified by applying  DC voltage signals (biasing) at the elements, which are typically PIN switch diodes and P-N varactor diodes. Switch diodes have been widely used in global or local binary-state scenarios for functional metasurfaces \cite{Mias:2007,Zhu:2010,Zhao:2013,Cui:2014,Li:2017,Yang:2016SciRep,Chen:2017,Georgiou:2018}. However, binary-state control still restricts the realizable functions as the impedance configurations along the metasurface are limited. Varactor diodes, on the other hand, provide continuous but mainly reactive control and, thus, the ability to tune only the imaginary part of the surface impedance. 

In Ref.~\cite{Liu:2019}, we presented a model metasurface with conceptual tunable integrated circuits (IC) incorporated in each unit cell to allow for full control over the complex surface impedance, i.e., independently and continuously variable resistance and capacitance. This \emph{full impedance control} enables one to spatially shape both the phase \emph{and} amplitude of the local reflection (or transmission) coefficient and allows maximum versatility in the realizable functionalities. More importantly, this concept can infuse ``intelligence'' in a metasurface programmatically controlled by a computer, the so called HyperSurface (HSF). \tblu{In addition, by interconnecting more advanced ICs (with more functions such as sensing the current besides providing the tunable loads) in all the unit-cells, forming a network, the vision of an \emph{intelligent metasurface fabric} can be realized, where each unit can sense the ambient conditions through the powerful IC, communicate information within the metasurface and with the main controller (computer), and perform its own programmable computing and actuation, with vast application within the emerging ``Internet of Things'' paradigm \cite{Liaskos:2015,Pitilakis:2018,Tasolamprou:2018,Tasolamprou:2019b,Petrou2018,Kouzapas2020,Saeed2018}.}

In this paper, we address the problem of designing a practical, multi-functional and reconfigurable metasurface for operation at microwave frequencies (5~GHz), taking fabrication constraints into account. We show, for the first time, control over the amplitude, wavefront, and polarization of the output wave in a single reconfigurable hardware platform.  The metasurface consists of square patches placed on a metal-backed dielectric layer. It includes an envisaged custom IC (or ``chip''), strategically placed behind the metal back plane, \tblu{allowing to impose independent complex loads to each patch}. We thoroughly present the design process of the multi-functional reconfigurable metasurface, from the perspective of the electromagnetic considerations and the implementation restrictions of the envisaged IC. We then evaluate the performance of the metasurface as a tunable perfect absorber, \tblu{which is the primarily targeted functionality}, and show successful operation at the design frequency for a range of incidence angles of both TE and TM polarizations. In addition, \tblu{we demonstrate its potential for wavefront manipulation functionality}, exemplified through anomalous reflection and polarization control capabilities, showcasing the multi-functional nature of the proposed system and its control over the amplitude, wavefront, and polarization properties of the output electromagnetic wave. Control over the frequency content of the impinging radiation could also be possible with the proposed linear metasurface, by using fast time modulation of the loads provided by the ICs. Additional steps towards the realization of this metasurface with a prime focus on the custom ICs implementation and practical aspects of the PCB design can be found in \cite{paper2b}.

The paper is organized as follows: in Section~\ref{sec:2:Evolution}, we present the evolution of the metasurface unit-cell design from the simple proof-of-concept version to a practically feasible prototype; dedicated subsections address different design aspects pertaining to metasurface topology and operation, basic structural decisions, {available range from the custom IC, fabrication and practical restrictions, and, finally, the simulation design strategy we have followed, relying on S-parameters of both lumped and Floquet ports.} In Section~\ref{sec:3:EMDesign}, we present the final design of the unit-cell of the \tblu{tunable perfect absorber} metasurface, backed by full-wave simulation results for performance evaluation {and parameter sensitivity}. In addition, we assess the performance for indicative beam steering and polarization conversion functionalities. Finally, Section~\ref{sec:4:Future} provides a summary and discusses future prospects and open challenges of such reconfigurable metasurface applications. 

\section{Conceptual design and practical considerations} 
\label{sec:2:Evolution}
Electromagnetic design of a locally tunable multi-functional metasurface, that can be readily manufactured with existing fabrication processes and at a reasonable cost, poses practical challenges that can only be met with multi-facet considerations and performance compromises. This section addresses the design considerations needed to select and develop the most appropriate topology, allowing as general reconfigurability as possible and leading to a realistic metasurface design. We take into account both electromagnetic response and practical considerations, as well as simulated dynamic load tunability of the IC, \tblu{while accommodating for the broadest possible functionalities. Readers familiar with unit-cell design methodology can skip this section and jump to the final design and performance results, in Section~\ref{sec:3:EMDesign}.}  

\subsection{Electromagnetic considerations for the design of versatile and fully reconfigurable metasurfaces}
\tblu{Probably the most natural choice for a topology of a tunable metasurface for microwave applications is the classical high-impedance surface formed by a subwavelength array of metal patches over a grounded dielectric layer~\cite{Sievenpiper:1999,Sievenpiper:2002,Sievenpiper:2003}. The patches can be either connected to the ground by metal vias (the so-called ``mushroom'' structure \cite{Sievenpiper:1999}) or left floating \cite{Tretyakov:2003,Luukkonen:2008}. The vias pins affect the oblique-incidence  performance for TM-polarized excitation \cite{Luukkonen:2009,Luukkonen:2009a}. This structure allows control over both amplitude and phase of the reflection coefficient by controlling the patch array only. It is very simple, compact, and can be manufactured using conventional printed circuit board (PCB) technology. }

\subsubsection{Uniform distribution of control elements \emph{versus} clustering}

\tblu{The simplest way to tune the metasurface response is to modify the gap capacitance between the patches, using both varactors~\cite{Sievenpiper:2002,Sievenpiper:2003,Mias:2007,Costa:2008,Zhao:2013,Lin:2013} and MEMS capacitors~\cite{Chicherin:2006}. }
For plane-wave illumination, the response of the metasurface is determined by the parallel-type resonance of the distributed capacitance of the patch array and the inductance due to the magnetic flux between the ground plane and the patch array \cite{Tretyakov:2003}. Thus, it appears reasonable to connect tuning elements (most commonly, varactors) into every gap between metal patches, and this topology was indeed used in  \cite{Sievenpiper:2002,Sievenpiper:2003,Chicherin:2006,Mias:2007,Costa:2008,Zhao:2013}, and other works. 

\tblu{This approach works well for surface-uniform biasing or for slowly-varying biasing (tunable phase-gradient metasurfaces).} However, there are applications where fast (on the wavelength scale) tunable variations of the surface properties are needed. For example, to realize anomalous reflection from normal incidence to near-grazing direction, so called metagratings \cite{Younes:2017,Epstein2019,Popov2020} offer arguably the most reasonable solution. To realize this regime, we need to configure the metasurface as a periodical array of electrically small elements with the period of the order of the wavelength. Moreover, to enable anomalous reflection into arbitrary directions, we need to have a possibility to configure the metasurface as a nonlocal, inhomogenizable  periodical array (with the period from $1.5\lambda$ to $3\lambda$) where each super-cell contains several small elements, all of them being different \cite{Diaz-Rubio:2017}. The conventional topologies of tuning the surface-averaged sheet reactance of the patch array are not optimal for realization of these more demanding functionalities, because the change of each control element strongly affects the collective response of the whole array of patches \cite{Diaz-Rubio:2017}. 

\begin{figure}[!t]
	\centerline{\includegraphics[width=85mm]{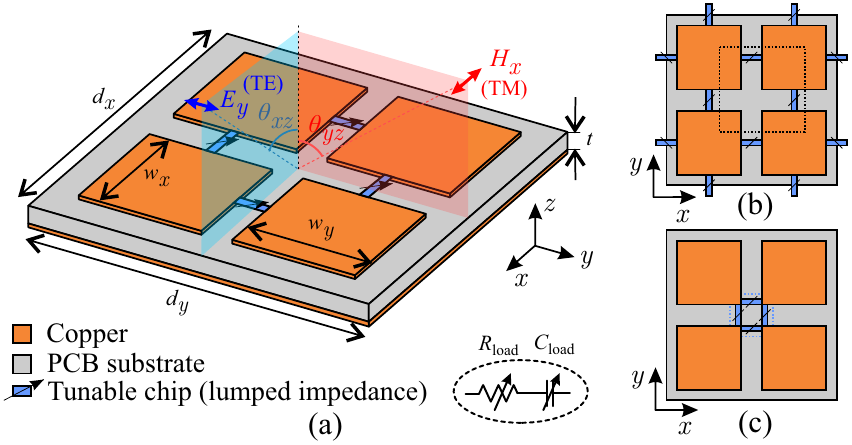}}
	\caption{(a)~Extension of the proof-of-concept $1\times2$-patch unit cell to an isotropic $2\times2$ cell. The new cell comprises four square patches on a thin metal-backed dielectric substrate. Its response is controlled by the ICs, modeled as complex lumped loads connecting each pair of patches. Variants of the isotropic unit cell with (b)~two ICs and halved width, as denoted by the dotted-line rectangle, (c)~a single IC incorporating all four connections, in the middle of a full-width cell.}
	\label{fig:1:PoCuc+2x2extension}
\end{figure}

\tblu{For this reason, it appears preferable to form the metasurface as an array of patch clusters with independent control. The simplest topology of a cluster consists of four patches with control units between them, as illustrated in Fig.~\ref{fig:1:PoCuc+2x2extension}(a). In this topology, the patches of the two neighboring clusters are coupled only through the gap capacitance, while in the usual, uniform layout, there is also the admittance of the control unit. Such a cluster can be viewed as one patch with tunable anisotropic sheet impedance. }Setting several clusters in the same or similar states, it is possible to tune the surface-averaged sheet impedance and realize phase-gradient metasurfaces or tunable absorbers. In the other extreme scenario, it is possible to connect four patches of one unit cell to the ground by different impedances, realizing one electrically small unit cell behaving as a tunable bianisotropic scatterer, while other clusters are tuned away from the resonance. This way it is possible to realize tunable diffraction gratings (metagratings \cite{Younes:2017}).  Finally, we note that the control elements can be active or time-varying, opening possibilities to create amplifying, non-Foster, and nonreciprocal reconfigurable metasurfaces. \tblu{The cluster layout provides the most general reconfigurability. Note that this patch layout is similar to that used in \cite{Lin:2013}, where only uniform bias has been considered. }

\subsubsection{Optimal patch shape}

\tblu{Next, we explain the reasons for selecting the square metal patches for varactor-tunable metasurfaces. In the literature on high-impedance surfaces and artificial magnetic conductors one can find suggestions of using patches of many different shapes, from simple Jerusalem crosses and various spiral shapes to space-filling curves, e.g. \cite{Simovski_Sochava:2003,Engheta:2004,Lee:2006}. Shapes other than simple squares are used when there is a need to reduce the thickness of the structure and/or the size of the unit cell. In all these solutions, the equivalent circuit of the patch array contains an additional inductance}. This inductance increases the total inductance of the parallel resonant circuit (the main inductance is due to the magnetic flux between patch array and the ground plane), bringing the resonant frequency down and allowing reduction of the substrate thickness. However, this additional inductance reduces the bandwidth of the resonator, because it is in series with the capacitive branch~\cite{Sievenpiper:2003,Costa:2009}. \tblu{Thus, if there is no need for reduction of the substrate thickness, it is clearly preferable to choose square patches with small gaps.} If there is a need of miniaturization, it is preferable to increase the effective capacitance of the patch array by using double layers of patches instead of replacing wide patches by thinner strips of any shape. 

\tblu{Another related issue is the isotropy of the surface. For most applications, it is desirable that the properties of the metasurface are isotropic in the metasurface plane. In this respect, the use of hexagonal patches is preferable, since the anisotropy of such arrays (assuming the same patch area) is considerably weaker than the square patch array.} However, this difference can be usually neglected for small arrays periods, and the square shape is still preferable due to a reduced number of control elements; in this work we adopt the square topology for the structure.  

\subsubsection{Optimal unit-cell size}

\tblu{In practice, it is desirable to use the smallest possible number of tunable unit cells, to minimize complexity and costs. This means that we should select the largest size of the unit cell, still allowing the target functions. If the target functionality is to control the reflection for plane-wave illuminations (e.g., tunable absorbers for varying frequency, incidence angle, and polarization), there is no need for control of the surface properties at the subwavelength scale. A periodical array with period smaller or equal to $\lambda/2$ will not create grating lobes, and we can configure all the unit cells identically, realizing an effectively uniform lossy boundary, matched to free space for the desired frequency, incidence angle, and polarization. Thus, for this application the optimal unit cell size equals $\lambda/2$ at the highest operational frequency.}

\tblu{Another important functionality is anomalous reflection with moderate deflection angles. In this case, the array should be configured as a phase-gradient metasurface, which is the conventional phased-array antenna (reflectarray) \cite{Pozar:1997}.} As is well-known, also in this case the use of $\lambda/2$-size array elements is appropriate. However, in this case the array is not uniform, and there is some scattering into parasitic propagating Floquet modes. Reducing the unit-cell size makes the effective (surface-averaged) response more uniform and reduces scattering  and improves the bandwidth \cite{Atef:2010,Pozar:1997}. However, improvements are not very dramatic, because for moderate deflection angles (for normal illumination, up to about $30^\circ$ to $40^\circ$) the parasitic scattering is small \cite{Diaz-Rubio:2017}, with only a few percent loss in power efficiency.

\tblu{More severe limitations need to be considered if the goal is to enable arbitrary shaping of reflected waves in the far field. In general, two different configurations are needed}: diffraction grating (metagrating) and  nonlocal metasurfaces. In the first case, the period is of the order of the wavelength and we need to ``activate'' only one or two elements, while the rest are set far from the resonance. In the second case, the period is larger than $\lambda$ but still comparable to it. As demonstrated in \cite{Diaz-Rubio:2017}, in this case we need about $8$ to $10$ unit cells per period, and all the cells are in general configured differently to enable efficient anomalous reflection to large angles. \tblu{To enable both these regimes, it is appropriate to select the unit-cell size not larger than about $\lambda/5$. Together with the clustering discussed above, selection of this size enables the most general reconfigurability of metasurface reflectors.}

\subsubsection{Choice of the substrate permittivity and thickness} 

\tblu{As is well-known from the theory of high-impedance surfaces, higher permittivity improves the stability of the resonant frequency with respect to the incidence angle and allows reduction of the substrate thickness, but decreases the bandwidth~\cite{Costa:2009}. For reconfigurable surfaces, the angular response can be adjusted by tuning the unit cells, while wider frequency bandwidth is desirable for robust operation.} In addition, there is also the practical consideration of availability of cheap and fabrication-friendly low-loss substrates. In view of these consideration, we select moderate-permittivity low-loss dielectric substrates. The thickness of the substrate mainly determines the effective inductance of the equivalent parallel resonant circuit which models the high-impedance circuit \cite{Sievenpiper:1999,Tretyakov:2003}. Higher inductance means higher bandwidth, meaning that thicker substrates are preferable. Naturally, the thickness should be still considerably smaller than the wavelength.

\subsubsection{Proof-of-concept study}

These general design considerations have been recently validated for  plane-wave illuminations in two orthogonal planes, parallel to the principal axes of an array of  square patches~\cite{Liu:2019}. The unit cell of this metasurface is schematically shown in  Fig.~\ref{fig:1:PoCuc+2x2extension}(a). In these illumination scenarios it is enough to consider only pairs of loaded square patches, because there is no current over the other two loads. The load impedance was assumed to be complex, with variable capacitance and resistance. The studied range of capacitance variations was $1$~pF to $5$~pF, corresponding to the parameters of commercially available chip varactors for the target frequency range (about 5~GHz) and similar to the ranges of varactors used in recent experimental studies of tunable high-impedance surfaces  \cite{Costa:2008,Lin:2013}. The tunable resistance was assumed to be varying from 0 to 5~Ohm, again in harmony with the parameters of commercial chip varistors, and allowing comparison with ideal, lossless structures. 

The most critical issue was to validate the hypothesis that it is possible to realize  near-perfect anomalous reflectors, not limited to any angular sectors, by tuning only load reactances of unit cells, while their sizes and shapes remain fixed and the same for all cells. The only known earlier realization of this functionality relied on careful optimization of \emph{sizes} of subwavelength patches forming super-cells of a periodical lattice \cite{Diaz-Rubio:2017}. The results presented in \cite{Liu:2019} have shown  that comparable performance can be achieved also using the proposed topology of fixed-size arrays of patch clusters with variable reactive loads, with practically realistic values of load capacitances. The study of tunable absorber configurations has shown that the same metasurface can be indeed configured  to the regime of full absorption of incident plane waves of both TE and TM polarizations, as depicted in Fig.~\ref{fig:1:PoCuc+2x2extension}(a), in a wide range of incidence angles.

\subsection{From the proof-of-concept unit cell to fully reconfigurable implementable unit cell design}

\tblu{For practical implementation, we start from the structure in \cite{Liu:2019}, which is only a proof-of-concept design and there is still a long way toward industrial realization.} For example, some open questions include: How should the chip be connected to the patches and how would the connectors affect the performance of the metasurface? Is the design compatible with fabrication technologies? How can the capacitive and resistive loads be optimally realized in practice? In this and the following subsections, we discuss all these practical issues and, based on these considerations, we propose a practically-realizable reconfigurable metasurface. 

\subsubsection{One chip instead of four chips per cluster}

The cluster topology shown in Fig.~\ref{fig:1:PoCuc+2x2extension}(a), with four chips connecting the patches, is not cost effective or practical to populate. \tblu{In addition, more complex routing lines are required when communications between the chips are demanded for programmability. Moreover, when the unit cell size is in the 10~mm order ($\lambda/6$ at 5~GHz), squeezing four chips inside such a small area demands advanced fabrication technologies, which again raise the cost. A simple and effective solution is to use just one chip inside each unit cell, as shown in Fig.~\ref{fig:1:PoCuc+2x2extension}(c). To enable the same functionalities, one chip is now essentially supporting at least four $RC$ pairs. With this simplification and the change of the $RC$ connection from the middle of the patch edge to the patch corner, the metasurface can still support the desired functionalities.} In addition, we note that in this way we also select the best position for the load connection, i.e., the patch corner, because that is where the surface charge density maximum across the patch is located for illumination in both primary incidence planes of the structure shown in Fig.~\ref{fig:1:PoCuc+2x2extension}(a).

\subsubsection{Load connection topology inside the chip} 

\tblu{Another issue raised in the ``one chip per cell'' approach is the choice of the patch-load-patch connecting topology, as well as the number of loads inside the chip. For practical reasons related to unavoidable parasitics inside the chip, as detailed in \cite{paper2b}, an X-shaped connection was finally adopted with four independent loads connecting the corner of each patch to the common ground pin in the center of the chip. This load topology can readily implement the absorber functionality by setting all loads to the same $RC$ values, and also accommodate other functionalities, e.g., anomalous reflection and polarization conversion (with different $RC$ values). }

\subsubsection{Out of plane chip placement}

\begin{figure}[!t]
	\centerline{\includegraphics[width=85mm]{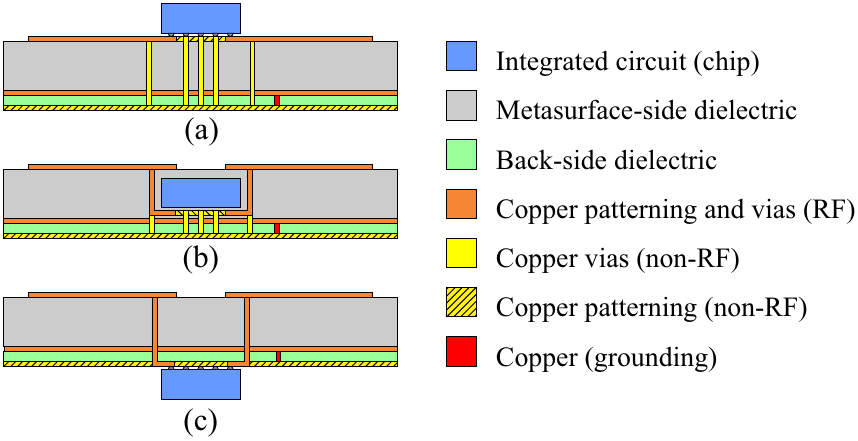}}
	\caption{Investigating the optimal vertical position of the chip, with respect to the number of vias and metallization layers required. Vertical cross-sections of possible designs where the chip is (a) on the top layer, (b) embedded inside the dielectric, (c) behind the backplane.}
	\label{fig:1:Fabrication}
\end{figure}

\tblu{A key decision that should be made is the placement of the chip in the vertical sense. Three options were examined: (a) on the top patch layer, (b) embedded inside the substrate, and (c) below the ground plane, as schematically shown in Fig.~\ref{fig:1:Fabrication}. Option (a) was abandoned because the the chip would interfere with the incident radiation and disrupting the performance; additionally, a large number of through vias (TVs) together with traces and landing pads (as TVs cannot be fabricated directly below the chip) would be required to pass communication and powering lines behind the backplane, and these vias/traces/pads would complicate and degrade the electromagnetic response, especially for TM polarization. Option (b) was also abandoned because embedding the chips involves costly non-standard fabrication processes; moreover, it excludes probing and servicing malfunctioning chips post-fabrication, and also requires a large number of blind vias. The optimal option for chip placement is behind the backplane, with only four TVs penetrating it to connect to the back-side of each patch, as shown in Fig.~\ref{fig:1:Fabrication}(c). From the electromagnetic design point of view, option (c) ensures all the auxiliary chip wiring does not affect wave propagation while the metasurface properties can be reconfigured. }

\tblu{Finally, a last issue is the position of the four TVs connecting the chip RF terminals to the patches. While in mushroom structures the vias are located at the center of the patches, here we stress that, reconfigurable metasurface, the optimal position for vias is at the patch corner which is the closest to the chip.} The corner of the patch is where the surface current density is at its maximum for illumination in both primary incidence planes and therefore the impedance introduced by the chip can more widely control the metasurface response.

\subsection{Integrated circuit design }

\begin{figure}[!t]
    \centerline{\includegraphics[width=85mm]{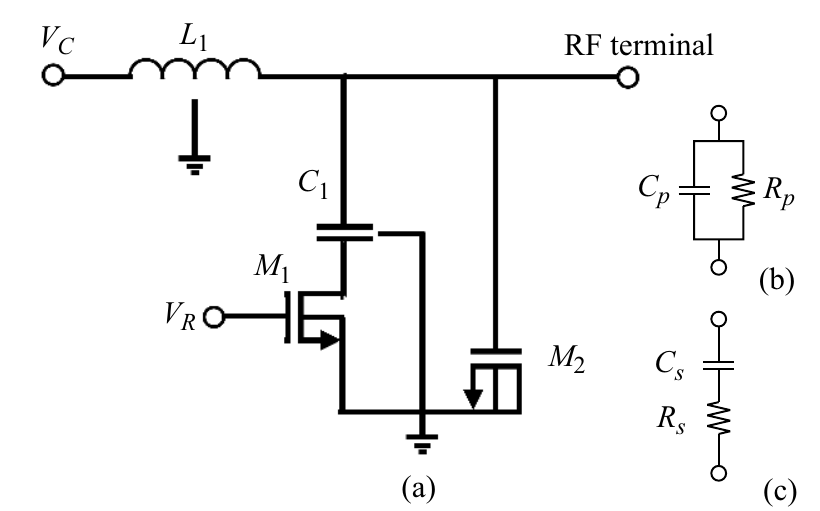}}
    \caption{(a) Simplified metasurface loading element circuit. Equivalent (b) parallel and (c) series $RC$ circuit representation of the metasurface loading element, between the ground and the RF terminal/port. }
    \label{fig:1:ChipDesign}
\end{figure}

In this section, the custom IC design of the metasurface loading elements ($R$ and $C$) is presented. \tblu{The architecture of the ICs that enables the independent control of the four complex loads was presented in \cite{paper2b}, it consists of an asynchronous circuit controller \cite{Petrou2018}, eight digital to analog (D/A) converters and four loading elements.} The IC chip aims to satisfy metasurface performance, cost and power constraints. It utilizes MOSFETs to implement both the variable capacitance (varactor) $C$ and variable resistance (varistor) $R$. The metasurface loading element simplified circuit is shown in Fig.~\ref{fig:1:ChipDesign}(a). In this schematic, the varistor $M_1$ is adjusted trough a gate voltage ${V_R}$ (biasing voltage) and the varactor $M_2$ is adjusted through biasing voltage ${V_C}$. The RF-choke inductor $L_1$ is used to block the RF signal from shorting on the low impedance node ${V_C}$. Similarly, the DC-block capacitor $C_1$ is used to ensure the ${V_C}$ voltage shorting to ground through $M_1$. The circuit forms a parallel connection $RC$ load which, at its steady-state, will draw negligible current due to gate leakage. \tblu{The design methodology for the lumped elements and all the actual on-chip circuit implementations of all the custom IC components (varistor, varactor, RF-choke inductor, DC-block capacitor) can be found in \cite{paper2b} and references therein.} \tblu{We stress, once more, that our design relies on a custom IC to implement the varistor and varactor functionality, and not on discrete commercially available surface-mount devices.} 

\begin{figure}[!t]
    \centerline{\includegraphics[width=85mm]{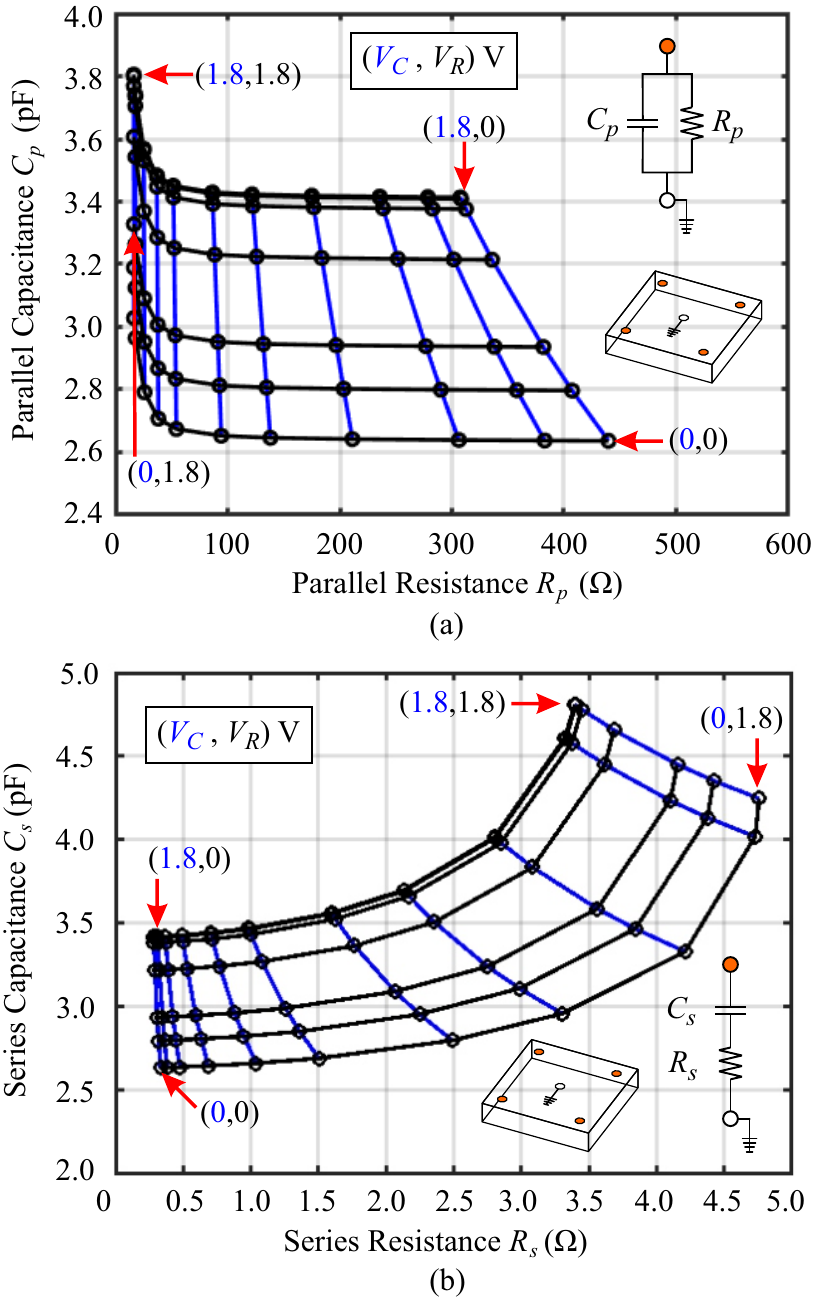}}
    \caption{Equivalent resistance and capacitance range for (a) parallel and (b) series representation\tblu{, at 5 GHz. The corresponding varistor and varactor voltage pairs are also annotated}. In the insets, the orange-filled and empty dots represent the RF terminal and the ground of Fig.~\ref{fig:1:ChipDesign}, respectively.}
    \label{fig:2:ChipDesign}
\end{figure}

\tblu{In the remainder of this work, we only consider the complex impedance values that are supplied by the custom IC to load the metasurface patches. The available complex impedance range, in terms of $RC$ values}, is shown in Fig.~\ref{fig:2:ChipDesign}, and was obtained by schematic simulations in Cadence Virtuoso. A 180~nm technology process design kit was used in S-parameter simulation. The obtained S-parameters were converted into Z-parameters and in turn converted to an equivalent parallel or series $RC$ load, as shown in Fig.~\ref{fig:1:ChipDesign}(b) or (c), respectively. In the Cadence simulations, the supplied voltages ${V_R}$ and ${V_C}$ are gradually increased from zero to IC’s supply voltage \tblu{(1.8 V)}. Each combination of ${V_R}$ and ${V_C}$ voltages corresponds to an implemented $RC$ value\tblu{; the limit $RC$ values, for the corresponding limit $({V_C},{V_R})$ pairs, are annotated on Fig.~\ref{fig:2:ChipDesign} with red arrows}. This defines an area with the achievable $RC$ values of the metasurface loading element. The values can be converted from parallel to series connection, and vice versa, using the following formulas
\begin{subequations}\label{eq:RC_P2S}
    \begin{align}
    R_s &= R_p(1 + Q^2)^{-1}, \label{eq:Rs}\\
    C_s &= C_p(1 + Q^{-2}), \label{eq:Cs}
    \end{align}
\end{subequations}
where $Q = \omega R_p C_p = 1/\omega R_s C_s$ is the quality factor of the circuit. In the remainder of the text, we will use the equivalent series $RC$ representation, unless otherwise mentioned.

\subsection{Fabrication and practical restrictions related to EM operation}

Having outlined the basic considerations that have led us to the approximate dimensioning of the metasurface unit cell in the lateral (cell and patch widths) and vertical (dielectric thickness) directions, and implementation rules for the custom IC, additional manufacturing limitations have been taken into account in the design. 

\begin{figure}[!t]
    \centerline{\includegraphics[width=85mm]{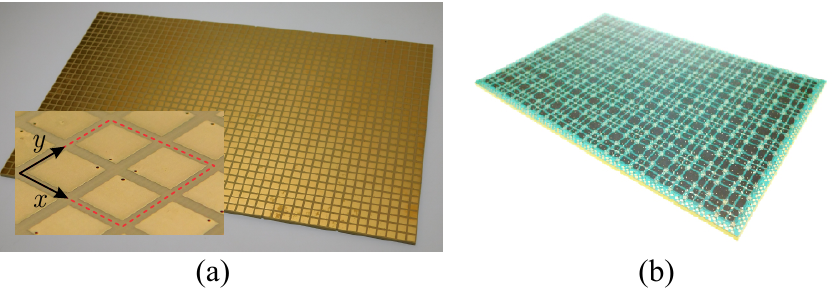}}
    \caption{ Photographs of the fabricated metasurface PCB, with a net area of 6.2''~$\times$~9.4'' corresponding to $24\times16=384$ unit cells. (a) Perspective view of the front side/top layer with Ni/Au metallization; inset shows a close-up view of the unit cell where the four electroplated (hollow) through vias can be seen. (b) Perspective view of the bottom side, where the custom ICs will be assembled.}
    \label{fig:1:IZMphotos}
\end{figure}

An important design aspect is manufacturability of the through vias (TVs) that will connect the IC RF terminals (on the back side of the metasurface) to the patches (on the front side of the metasurface). The available technology for TVs requires for a maximum 1:8 aspect ratio, i.e., the thickness of the material stack for the mechanical drilling of TVs must not exceed 8 times the diameter of the via, so as to achieve sufficient copper via electroplating. In contrast, blind vias (BV) require a 1:1 aspect ratio, leading to relatively thicker cylinders. Simulations and fabrication concepts have recommended an approximate thickness of the dielectric stack penetrated by the TVs to be 2.4~mm, and therefore a TV diameter of 0.3~mm was chosen. It was numerically verified that increasing the diameter of the TVs leads to small increase in the  resonance frequency, without degrading its quality factor (the amplitude of the reflection coefficient remains $-30$~dB or lower). For example, for TV diameter increase from 270 to 300~$\mu$m, the resonance frequency increases by roughly 70 MHz, which can be compensated by slightly increasing the capacitance introduced at the chip terminals, well within the tuning capabilities of the chip. Finally, it should be noted that all vias are copper electroplated so that they are essentially hollow metallic cylinders, inset of Fig.~\ref{fig:1:IZMphotos}(a); although our EM simulations were conducted assuming solid-copper vias, for the sake of simplicity, additional numerical simulations have confirmed that the two models are equivalent owing to the small skin depth ($\sim1~\mu$m at 5 GHz). 

The next practical considerations concerned the dielectric material choices to be used above and below the metal backplane. Initial fabricated prototypes used high-frequency graded Rogers RT/duroid 5880 dielectric ($\varepsilon_r=2.20$, $\tan\delta=0.0009$) on the top side, and regular FR4 dielectric (Hitachi MCL-E-679FGB) on the back side, where low loss is not as important to minimize cost. Fabrication of the initial 3-layer substrates, e.g., as in Fig.~\ref{fig:1:Fabrication}(c), resulted in significant warping (bending) of the printed circuit board after lamination of all build-up layers, even for small metasurface substrates, e.g., 10 cm by 10 cm, due to mismatch of the thermal-expansion coefficients of the two dissimilar dielectrics. For this reason, subsequent designs used the same dielectric substrate type, namely Panasonic Megtron7N dielectric ($\varepsilon_r=3.35$, $\tan\delta=0.0020$), with laminates and prepreg layers symmetrically positioned above and below the backplane, thus minimizing warping. Given the fixed dielectric thicknesses available, a stack of Megtron7N laminates and prepregs has been symmetrically laid up to achieve the total thickness required and ensure that warpage has been diminished; indeed, the homogeneous nature of prepregs and laminates used has yielded minimum warpage. Figure~\ref{fig:1:IZMphotos} provides a view of the demonstrator metasurface substrate made of Megtron7N PCB materials for the above mentioned studies. It has been 2.4~mm thick, 160~mm~$\times$~240~mm (6.2”~$\times$~9.4”) in size and has been produced in large industrial panels of 457~mm~$\times$~610~mm format.

\begin{figure}[!t]
    \centerline{\includegraphics[width=85mm]{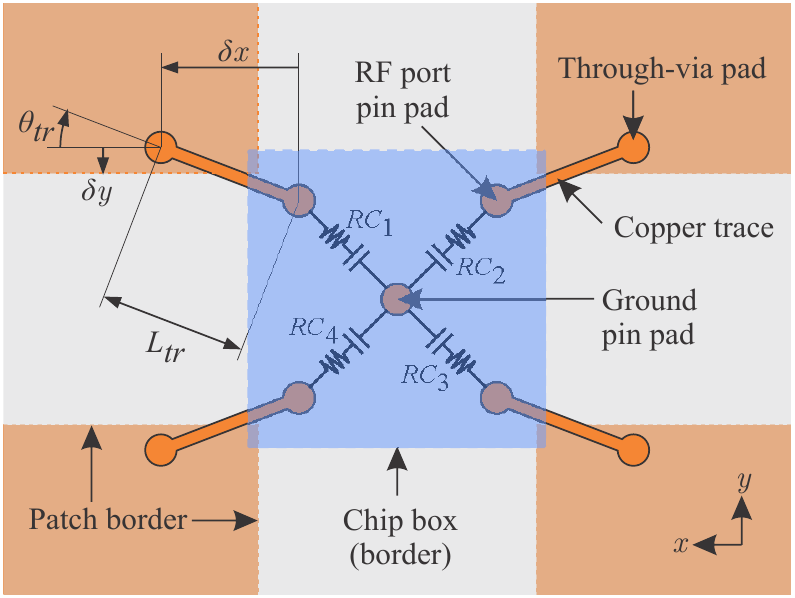}}
    \caption{View of the back side of the unit-cell, depicting the lateral offsets of the through via (TV) placement imposed by fabrication limitations. Parameters $\delta x$ and $\delta y$ denote the minimum lateral distance of the TV pad to the corner chip-pin pad and to the inner edge of the patch, respectively. \tblu{The four lumped $RC$ elements that load the patches through the TVs can be independently controlled.}}
    \label{fig:1:ViaOffsets}
\end{figure}

As discussed in the previous subsections, the optimal position to connect each pair of patches with lumped loads is at their corners; this can be intuitively understood by the increased surface current density at such positions which allows for wider tunability, and by the minimization of the inductance and resistance associated with short copper paths. However, in a realistic structure, the TVs that connect the patches to the IC (modeled by lumped loads) cannot be placed very close to the patch corner, for two fabrication-related reasons: firstly, because drilling of the TVs can only take place outside the chip footprint and, secondly, because the landing pads for the vias (drilled from the back to the front) are two times the via diameter and, thus, cannot be placed too close to the patch edge. The minimum values of these two offsets were chosen as $\delta x=0.4$~mm and $\delta y=0.2$~mm, respectively, and are depicted in Fig.~\ref{fig:1:ViaOffsets}. Please note that the resulting unit cell exhibits structural anisotropy, enforcing only mirror symmetry and not four-fold rotational symmetry, meaning that the metasurface will behave differently for the two normally impinging polarizations ($x$- and $y$-polarized). This compromise arose from limitations stemming from the $RC$ values due to the physics of real chips and the size of the patches, and will be discussed in detail in Section~\ref{sec:3:EMDesign}.

Concluding this part with restrictions related to the fabrication, firstly, it was verified that laterally offsetting the BV connecting the ground pin of the chip (which is the one in its middle) to the metasurface ground plane has no effect on TE polarized absorber performance, our main target. Therefore, in the simulations performed in this work a single BV was placed in the center of the chip footprint. However, in the fabricated device \cite{paper2b}, a copper trace brings this pin outside the chip footprint where the BV will short it to the metasurface ground plane, so that the chip terminals and the patches have a common reference ground. The length and path of this BV trace is expected to influence oblique TM performance due to mutual coupling with the RF terminal traces because of the current flowing through the BV (which is zero for TE polarization). Finally, in order to model the device as thoroughly as possible, one would need to quantify the effect of the solder balls used to electrically connect the RF terminals of the chip to their respective pads on the metallization layer.

\subsection{S-parameter modeling and simulation}

\begin{figure}[!t]
    \centerline{\includegraphics[width=75mm]{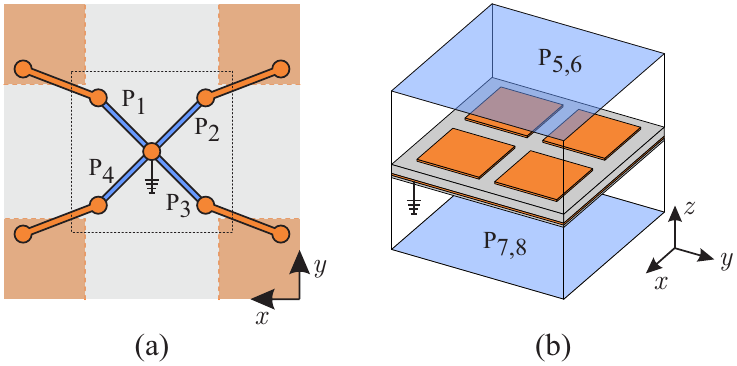}}
    \caption{Modeling of the unit cell using combination of lumped and Floquet S-parameter ports: (a) Back side of the unit cell where the four lumped ports, $P_{1,2,3,4}$, represent the RF ports (terminals) of the IC\tblu{, individually addressed similar to the $RC_{1,2,3,4}$ of Fig.~\ref{fig:1:ViaOffsets}}. (b) Perspective view of a subwavelength unit cell depicting the four Floquet ports, $P_{5,6}$ corresponding to TE and TM polarized ports on the front (top) side, and $P_{7,8}$ similarly on the back (bottom) side. Floquet ports are defined on planes perpendicular to the $z$-axis, and for a given incident plane wave direction.}
    \label{fig:1:CoSimulationStrategy}
\end{figure}

In order to efficiently model and simulate this multi-parametric EM problem, we rely on S-parameters, enabled by the metallic backplane, which effectively decouples the metasurface- and the chip-side designs. Thus, using the backplane as a common reference ground, we can model the chip with four lumped ports, Fig.~\ref{fig:1:CoSimulationStrategy}(a), and the incident and scattered plane waves with Floquet ports, Fig.~\ref{fig:1:CoSimulationStrategy}(b). The lumped ports are 50~Ohm-referenced and fed by Touchstone data (complex spectra) depending on the IC architecture, i.e, ranging from a single generalized four-port network (most complicated case) to four identical decoupled $RC$ loads (simplest case). The Floquet ports are defined for a given plane wave direction, i.e., a $(\theta,\phi)$ pair for incoming waves, for both polarizations (TE/TM), and are placed at the front and back side of the unit cell, at planes perpendicular to the $z$-axis; periodic Bloch-Floquet conditions are applied to the remaining four boundaries enclosing the unit cell, emulating infinite periodicity. Please note that in the case of uniformly configured chips we use a total of four Floquet ports: as the unit cell is subwavelength only the zero-order (specular) diffraction  order is propagating but we generally need to consider both linear orthogonal polarizations at both sides. In the case of inhomogeneous unit cell arrangement, where a supercell with periodicity exceeding the free-space wavelength is formed, such as the anomalous reflection scenario in Section~\ref{sec:sub:Steering}, the number of Floquet ports increases to accommodate the other propagating diffraction orders.

{For the perfect absorber functionality all unit cells of the metasurface must be identically configured; also, for our metal-backed weakly anisotropic structure, we can safely assume that transmission and cross-polarized reflection coefficients will be practically zero, since the structural anisotropy was found too weak to induce polarization conversion. For these reasons, we model a single unit cell and we seek the IC configuration that will lead to a minimization of the reflection coefficient $|r|$ for a given polarization, i.e., a minimization of the $|S_{55}|$ or $|S_{66}|$ scattering parameter for TE and TM incidence, respectively, using the port-number convention of Fig.~\ref{fig:1:CoSimulationStrategy}. In the general case, the power absorption coefficient when exciting Floquet port $i$ is $A_i = 1-\sum_{j}|S_{ji}|^2$, where $j$ runs through the Floquet ports (5 to 8).}

As stated in Section~\ref{sec:2:Evolution}-D, we have intentionally designed the chip with its four RF ports decoupled, so that they can be directly replaced by lumped loads, with their $RC$ values falling inside the prescribed IC map, Fig.~\ref{fig:2:ChipDesign}. This allows us to rapidly find the chip configuration, in terms of $RC$ pair, that leads to perfect absorption at a given frequency and polarization: For a given incident plane wave direction, we numerically evaluate the 8-port S-parameters of the unit cell with full-3D EM simulation; then, we attach a prescribed $RC$ load to all four lumped ports (passed as 1-port Touchstone data) and, finally, we extract the scattered spectra at the remaining four Floquet ports, corresponding to the two polarizations (TE and TM) and to the two sides of the unit cell (top/back side, i.e., reflected/transmitted). Iterating over the $RC$ loads using an optimizer, we can rapidly identify the optimal configuration for perfect absorption at a given frequency and polarization. We have conﬁrmed that within the given RC range from the chip design, this method works perfectly as the load range is limited. However, note that changing a geometric/EM parameter of the structure and/or the incident plane wave direction, e.g., for oblique incidence absorption, requires numerical recalculation of the 8-port network S-parameters with a new full-3D EM simulation, which is computationally intensive (time consuming) due to the fine geometrical features. {These simulations were conducted in CST Microwave Studio, using the 3D frequency domain solver with broadband sweep and the Design Studio environment for the 8-port S-parameter tasks.}

\section{Electromagnetic design of the complete unit cell and the multi-function performance} 
\label{sec:3:EMDesign}
In this Section we present the metasurface unit cell design, taking into account all the aspects and restrictions discussed in Section~\ref{sec:2:Evolution}. Prime focus is given on wide-angle tunable perfect absorption functionality by thoroughly analyzing its performance; steering and polarization conversion functionalities are also assessed. \tblu{The different applications require different load settings within and between chips: (i) in perfect absorption (Section III.A), all four loads in the unit cell are set to the same value and all unit cells are collectively tuned (uniform metasurface); (ii) in anomalous reflection (Section III.B), all four loads in the unit cell are set to the same value, but the unit cells comprising the supercell are tuned differently; (iii) in polarization conversion (Section III.C), all cells are collectively configured but the loads in the unit cell are set to different values to break the four-fold symmetry of the unit cell. These settings apply for plane incoming/outgoing wavefronts and a planar (flat) metasurface. }

\subsection{Tunable Perfect Absorption} 

\begin{figure}[!t]
    \centerline{\includegraphics[width=85mm]{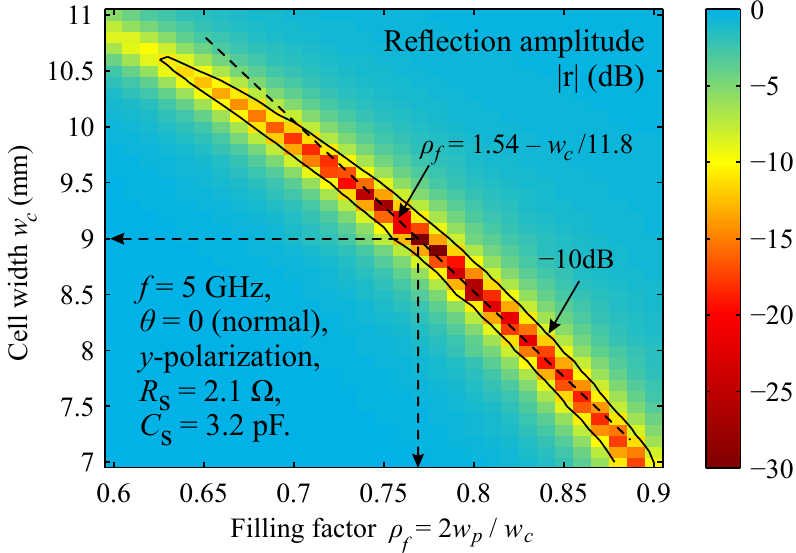}}
    \caption{Design map of the unit cell, depicting the reflection amplitude of $y$-polarized normally incident plane wave at 5~GHz, assuming the chip is configured to deliver the equivalent lumped loads $R_s=2.1$~Ohm and $C_s=3.2$~pF at its ports. This map allows the extraction of the optimal unit-cell and patch width, $w_c$ and $w_p=\rho_f w_c/2$, respectively, with all the remaining structure parameters defined in Section~\ref{sec:2:Evolution}-D. }
    \label{fig:2:DesignRules}
\end{figure}

\begin{figure}[!t]
    \centerline{\includegraphics[width=85mm]{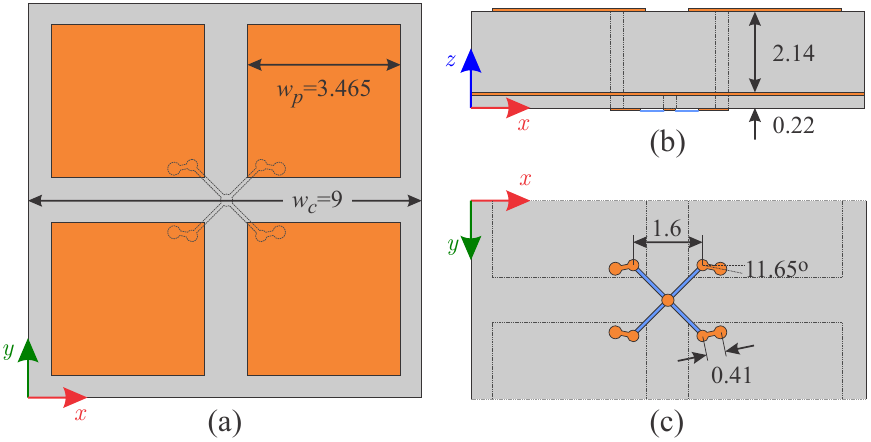}}
    \caption{Final unit-cell design and dimensions in millimeters. (a) Top view, showing the cell and patch widths. (b) Side view cross-section, showing the dielectric thickness of the top and bottom dielectrics. (c) Bottom view, showing: the X-connection shape of the four lumped ports to the ground, the locations of the through-vias connecting the chip RF terminals to the patches, and the respective traces.}
    \label{fig:2:UnitCellDesignSchems}
\end{figure}

The main functionality implemented by our reconfigurable metasurface device is tunable perfect absorption (PA) of plane waves, which can impinge from a large cone of incidence directions and both polarizations. As discussed, for the IC-loaded unit cell to achieve PA at a given frequency, polarization, and direction of incidence, we must properly tune both the resistive and reactive (capacitive) part of the equivalent lumped loads inside the IC. Given the tuning ranges of the varistor and varactors in the chip design, Section~\ref{sec:2:Evolution}.C, the optimization of the free structural parameters of the unit cell design is performed. The reference optimization case is PA at 5~GHz, normal $y$-polarized incidence, $R_s=2.1~\Omega$, and $C_s=3.2$~pF; these $RC$ values lie approximately in the middle of the chip range, thus allowing maximum tunability. The free structural parameters of the design are essentially the cell and patch width, $w_c$ and $w_p$, respectively, under the condition that the through vias are placed as close as possible to the inner patch corners, which is limited by the fabrication thresholds $\delta x$ and $\delta y$ discussed in Section~\ref{sec:2:Evolution}.D. In order to unveil the underlying design rules, one can parametrically calculate the reflection amplitude as a function of $w_c$ and the newly defined filling ratio, $\rho_f=2 w_p/w_c$, with the results presented in Fig.~\ref{fig:2:DesignRules}. Full-wave EM simulations of the unit cell are employed, conducted in CST Microwave Studio, using the frequency-domain solver with tetrahedral mesh. An approximately linear trendline emerges, showing that smaller cells require for proportionally larger patches to resonate for the same external load values. For our design, PA is achieved at the optimal values of $w_c=9$~mm and $\rho_f=0.77$, i.e., $w_p=3.465$~mm. For these parameter values, the length and angle of the metal traces connecting the RF terminals of the chip to the landing pads of the through vias defined in Fig.~\ref{fig:1:ViaOffsets} are $L_{tr}=0.41$~mm and $\theta_{tr}=-11.65^\circ$, respectively. The resulting unit cell and structural dimensions are summarized in Fig.~\ref{fig:2:UnitCellDesignSchems} in millimeters.

\begin{figure}[!t]
    \centerline{\includegraphics[width=75mm]{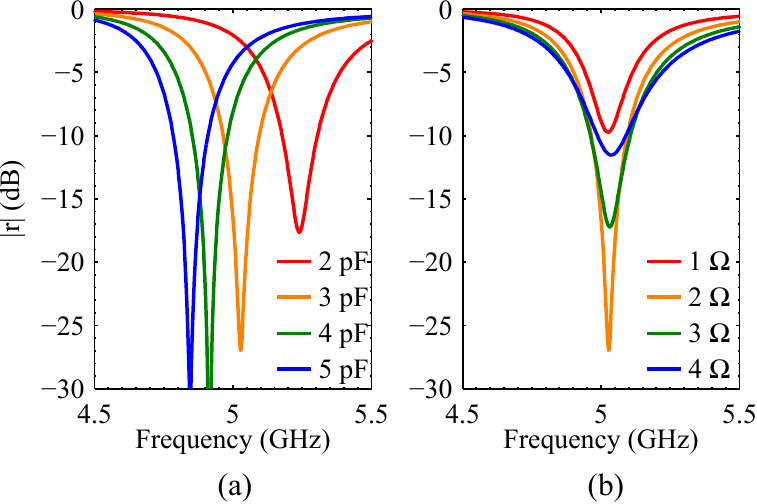}}
    \caption{Indicative spectra of the reflection amplitude for normal $y$-polarized incidence. (a) Series capacitance tuning affects the absorption resonance frequency for fixed $R_s=2~\Omega$, and (b) series resistance tuning affects the resonance quality factor, i.e., ``depth'', for fixed $C_s=3$~pF. }
    \label{fig:2:AbsorberSpectra}
\end{figure}

\begin{figure}[!t]
    \centerline{\includegraphics[width=70mm]{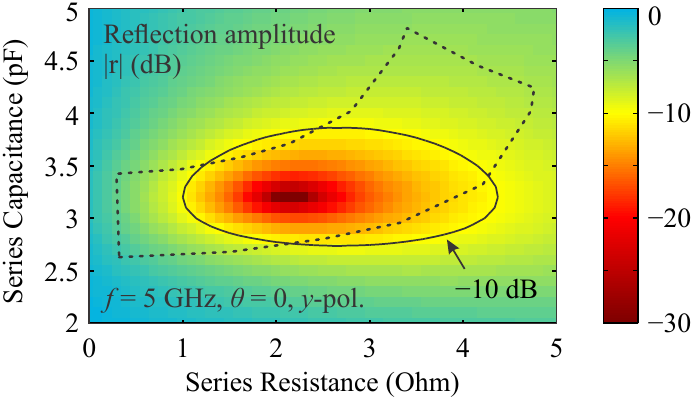}}
    \caption{Reflection amplitude for $y$-polarized normal incidence at 5~GHz, as a function of identical series $RC$ loads attached to all four ports. The dotted area is the $RC$ values allowed by the IC.}
    \label{fig:2:RCtolerance_5GHz_normal_yPol}
\end{figure}

Having completed the selection of all the structural dimensions, the performance for PA functionality is assessed. The amount of control imparted by resistance and capacitance tuning is depicted in Fig.~\ref{fig:2:AbsorberSpectra}, where $C$ and $R$ control the resonance frequency and quality factor (resonance ``depth''), respectively; these spectra are for normal $y$-polarized incidence. The resonance bandwidth is over 100~MHz (or 400~MHz) measured at $-10$~dB (or $-3$~dB). The respective tolerance values of $C$ and $R$ for near-perfect absorption at 5~GHz are quantified in terms of the achieved reflection amplitude, $|r|<-10$~dB, Fig.~\ref{fig:2:RCtolerance_5GHz_normal_yPol}. The attained range falls well within the varistor and varactor range of the chip thus allowing tuning and possible fabrication-deviation compensation.

\begin{figure}[!t]
    \centerline{\includegraphics[width=75mm]{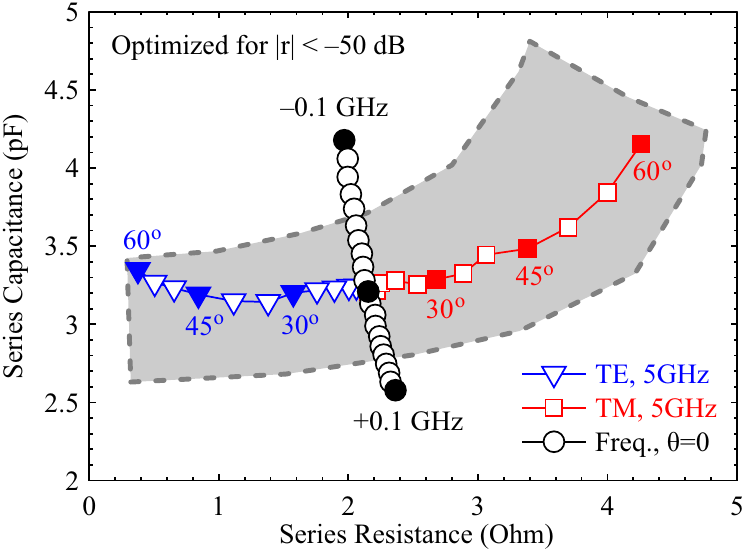}}
    \caption{Coverage for oblique incidence and frequency detuning for the tunable perfect absorption functionality of the final unit cell in terms of series $RC$ loads. Oblique incidence is at 5~GHz with angle steps of $5^\circ$; frequency detuning is in steps of 10~MHz. $y$-polarized incidence is assumed in all cases, i.e., the plane of incidecne is $xz$ and $yz$ for TE and TM case, respectively.}
    \label{fig:2:CoverageObliqueFreqTune}
\end{figure}

The last step in validating the performance of the absorber is to quantify the coverage range, in terms of oblique incidence angle and/or frequency detuning, allowed by the available $RC$ map of the IC. The S-parameter block modeling approach described in Section~\ref{sec:2:Evolution}.E is employed in order to efficiently calculate the optimal load impedance values for various combinations of incidence direction, polarization, and frequency. The results are presented in Fig.~\ref{fig:2:CoverageObliqueFreqTune}. Oblique incidence coverage is up to $60^\circ$ for both TE and TM polarization planes as defined in Fig.~\ref{fig:1:PoCuc+2x2extension}(a), i.e., with the electric field always primarily polarized along the $y$-axis, where the lateral distance between the through vias is smallest. Frequency tuning is limited to $\pm 50$~MHz, owing to the sharp absorption resonance and the limited capacitance range.

\subsubsection{Cross-polarized absorption performance}

\begin{figure}[!t]
    \centerline{\includegraphics[width=70mm]{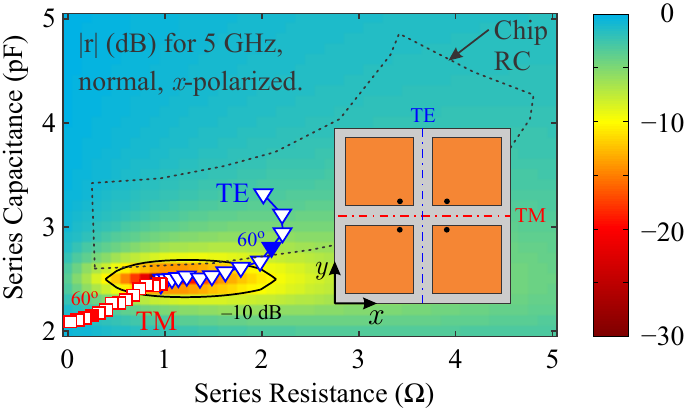}}
    \caption{Cross-polarized tunable absorption performance at 5~GHz. The colormap corresponds to reflection amplitude of $x$-polarized normal incidence as series $RC$ lumped loads are varied. Optimal $RC$ values for oblique incidence ($5^\circ$ steps) in TE and TM planes are superimposed, with blue triangles and red squares, respectively. The inset depicts how TE and TM planes are defined in the anisotropic unit cell for the cross-polarized case.}
    \label{fig:2:xPol_Performance}
\end{figure}

Since the targeted metasurface functionality was widely-tunable perfect absorption for at least one polarization, the unit cell design was allowed to be anisotropic, i.e., to exhibit mirror symmetry but not four-fold rotational symmetry, as depicted in Fig.~\ref{fig:2:UnitCellDesignSchems}(a). In this regard, $y$-polarization (i.e., $E_y$ in TE and $E_{yz}$ in TM) will outperform $x$-polarization, for perfect absorption on the given $RC$ range. We opted for this trade-off, in order to have at least one of the polarizations perfectly absorbed and with wide oblique incidence tunability; trying to accommodate both polarizations in the limited $RC$ range while enforcing four-fold symmetry (i.e., traces angle $\theta_{tr}=45^\circ$ and $L_{r}\leq\sqrt{2}dx$, as defined in Fig.~\ref{fig:1:ViaOffsets}) led to meager simulated performance for both polarizations and/or very limited oblique incidence coverage. Nevertheless, we assessed the cross-polarized absorption performance, with the results summarized in Fig.~\ref{fig:2:xPol_Performance}. As expected, very limited coverage is offered, e.g., only highly-oblique $x$-polarized (TE) incidence can be perfectly absorbed for the given $RC$ coverage, while suboptimal performance, e.g., reflection as high as $|r|\approx-10$~dB, can be attained for normal incidence.

\subsubsection{Double resonance at highly oblique incidence}

\begin{figure}[!t]
    \centerline{\includegraphics[width=85mm]{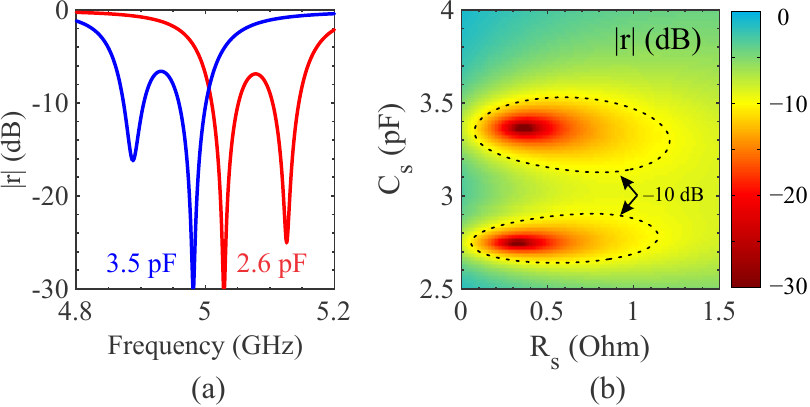}}
    \caption{Double absorption resonance arising for very oblique TE incidence angles. (a) Refection amplitude spectra for $\theta=60^\circ$, $y$-polarized, $R_s=0.35~\Omega$, and two different values of series capacitance $C_s$. (b) Reflection amplitude at 5~GHz, depicting the two optimal series-$RC$ configurations available.}
    \label{fig:2:DoubleDip}
\end{figure}

It must be noted that highly oblique incidence cases give rise to an interesting phenomenon of dual absorption resonance. For instance, in the $\theta=60^\circ$ TE $y$-polarized case, we can clearly see two minima of reflection spaced by approximately 100~MHz, Fig.~\ref{fig:2:DoubleDip}(a). Naturally, the unit cell $RC$ values can be tuned to optimize either of the two resonances for absorption at the desired frequency, as depicted in the tolerance map of Fig.~\ref{fig:2:DoubleDip}(b). The series resistance is approximately the same, 0.35~Ohm in this case, while the series capacitance can vary up to 1~pF. This double resonance gradually emerges for $\theta>45^\circ$, being more pronounced for $y$-polarized waves in the C-band (5~GHz) for our unit cell design. 

\subsection{Anomalous Reflection}\label{sec:sub:Steering} 

\begin{figure}[!t]
    \centerline{\includegraphics[width=85mm]{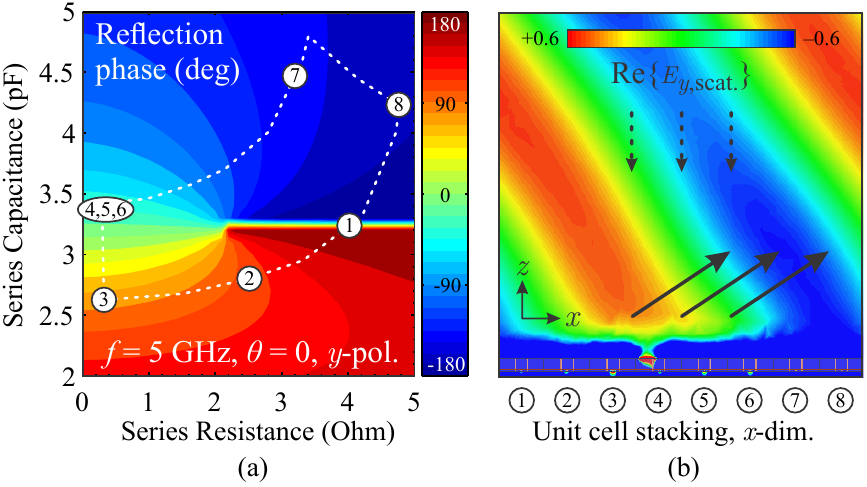}}
    \caption{(a) Reflection phase of homogeneous metasurface setting for 5~GHz, normal, $y$-polarized incidence; the white circles depict the series $RC$ settings for the eight unit cells, required for anomalous reflection function. (b) Scattered $E_y$ field pattern for anomalous reflection function.}
    \label{fig:2:ResultsSteering}
\end{figure}

The independent control over the $RC$ values in each unit cell enables us to perform wavefront manipulation operations, such as anomalous reflection. Although the reflection amplitude is \tblu{low} within the realizable $RC$ range by the chip (Fig.~\ref{fig:2:RCtolerance_5GHz_normal_yPol}), the coverage of the reflection phase is quite large, i.e., from $-170^\circ$ to $+170^\circ$, as shown in Fig.~\ref{fig:2:ResultsSteering}(a). As a result, anomalous reflection can be achieved if we lower the requirement on the reflection amplitude. Anomalous reflection is obtained by grouping several unit cells into a supercell and allowing the reflection toward the desired direction while forbidding the scattering to other directions by selectively promoting a single propagating diffraction order over the others. Utilizing different supercell sizes and relying on different diffraction orders, the anomalous reflection direction can be tuned with a quasi-continuous angle coverage \cite{Liu:2019}. To demonstrate the anomalous reflection effect, we consider a supercell made of $N=8$ unit cells, which will support anomalous reflection from normal incidence toward the $56.4^\circ$, according to $\theta_r = \arcsin{(\lambda_0/D)}$ \cite{Liu:2019,Tsilipakos:2018AOM}, where $D=Nd_x$ is the extent of the supercell. In this supercell configuration, which is different from the one in Fig.~\ref{fig:1:CoSimulationStrategy}, there are three ports, $a$, $b$, and $c$, corresponding to the three propagating diffraction orders $m=0,+1,-1$ (since $\lambda_0 < D < 2\lambda_0$). Note that transmission and polarization conversion is negligible and, thus, power collected by the corresponding ports is negligible. Subsequently, we perform an optimization while constraining the $RC$ values within the realizable range. Note that in order to speed up the optimization process one can use as an initial setting for the $RC$ values those dictated by a linear phase profile along the supercell, i.e., $\phi(x)=\phi_0-(2\pi/D)x$ \cite{Liu:2019}. 

The required $RC$ values for the 8 unit cells after optimization are shown as white dots in Fig.~\ref{fig:2:ResultsSteering}(a).   With  this configuration, the normally incident wave with polarization $E_y$ is reflected to $56.4^\circ$ ($m=+1$, port $b$) without prominent scattering to the specular direction ($m=0$, port $a$) or the $m=-1$ diffraction order ($-56.4^\circ$, port $c$), as the scattered $E_y$ field pattern shows in Fig.~\ref{fig:2:ResultsSteering}(b). The amplitude of the anomalous reflection coefficient is $|S_{ba}|=0.38$, which means that $14.4\%$ of the power is reflected toward the desired direction. Meanwhile, the amplitudes of the reflection coefficients to the other two directions are $|S_{aa}|=|S_{ca}|=0.05$. The absorbed power by the metasurface is given by $A_a\approx1-\sum_{n}|S_{na}|^2=85.1\%$, where $n$ runs through the propagating reflection diffraction orders $\{a,b,c\}$; an anomalous reflection efficiency of $\operatorname{Eff}_{ba}=|S_{ba}|^2/\sum_{n}|S_{na}|^2=96.7\%$ is calculated, meaning that $96.7\%$ of the scattered power goes to the desired direction. \tblu{The low reflection amplitude is primarily due to the custom IC imposing a limited $RC$ tuning range and, secondarily, due to fabrication restrictions; in principle, having a broader $RC$ range and less fabrication constraints would allow tuning the unit cell far from its absorption resonance thus allowing much higher reflection amplitude with ample phase span.} Finally, we emphasize that through the same principle of locally modifying the reflection phase by changing the $RC$ loads, we can achieve more wavefront manipulation operations, such as retroreflection and focusing~\cite{Liu:2019,Tsilipakos:2018AOM}.

\subsection{Polarization Conversion}\label{sec:sub:Polarization} 
Next, we exploit additional advantages of the proposed unit cell configuration. Organizing patches in groups of four and using a single chip to control them was judiciously chosen to save resources and, in principle, to lead to a \emph{geometrically} isotropic unit cell design that could have polarization independent response. On the other hand, however, this patch connectivity offers the distinct capability of \emph{electrically} breaking any rotational symmetry, thus leading to polarization conversion.

\begin{figure}
    \centerline{\includegraphics[width=85mm]{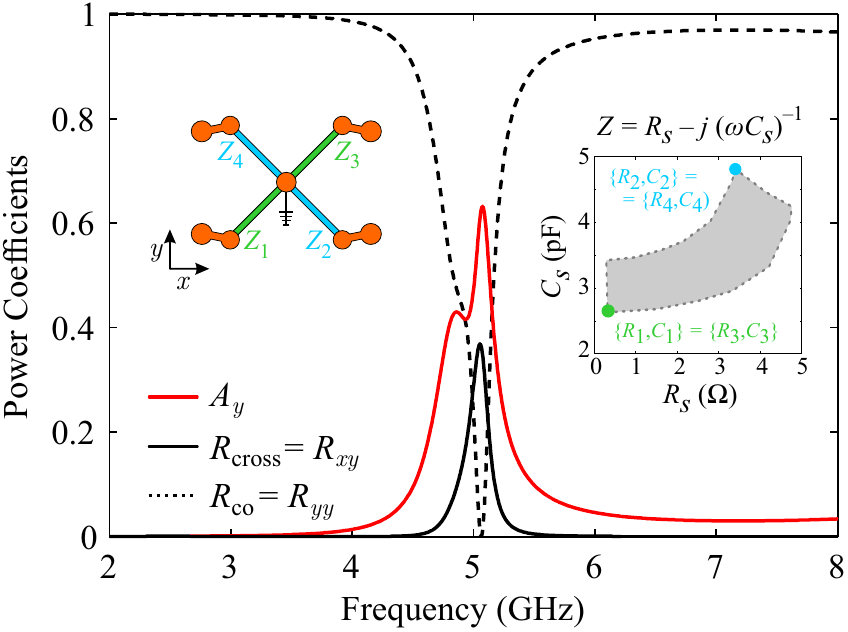}}
    \caption{\label{fig:ResultsPolarization}Linear polarization conversion by electrically breaking the four-fold rotational symmetry using loads $\{R_1,C_1\}=\{R_3,C_3\}=\{0.35~\Omega, 2.6~\mathrm{pF}\}$ and $\{R_2,C_2\}=\{R_4,C_4\}=\{3.4~\Omega, 4.8~\mathrm{pF}\}$ to emulate a 45$^\circ$ tilted cut wire. Complete polarization conversion to the orthogonal state is observed. The cross-polarized power coefficient is $37\%$ and absorption amounts to $61\%$. The left and right insets depict the load connectivity and the available $RC$ range, respectively. 
    }
\end{figure}

Here, as a demonstration of the opportunities for polarization control, we present linear polarization conversion to the orthogonal state. To this end, we appoint asymmetric load values to the four patches. More specifically, we use the complex loads $\{R_1,C_1\}=\{R_3,C_3\}$ and $\{R_2,C_2\}=\{R_4,C_4\}$ to electrically emulate the geometric structure of a 45$^\circ$ tilted cut wire, which has been shown in the literature to result in efficient linear polarization conversion \cite{Grady:2013}. We find that the optimal series $RC$ values are $\{R_1,C_1\}=\{R_3,C_3\}=\{0.35~\Omega, 2.6~\mathrm{pF}\}$ and $\{R_2,C_2\}=\{R_4,C_4\}=\{3.4~\Omega, 4.8~\mathrm{pF}\}$, since they allow for the maximum disparity regarding the capacitance values between on- and off-diagonal elements, while retaining the total resistance at the lowest possible levels. (Note that for the these $RC$ values the reactances dominate over the resistances at the neighborhood of 5~GHz.) We succeed in getting perfect polarization conversion to the orthogonal state, i.e., $R_\mathrm{co}=R_{yy}=0$ at 5.05~GHz, as depicted in Fig.~\ref{fig:ResultsPolarization}. However, having used resistance values as high as 3.4~$\Omega$ for the off-diagonal elements the absorption is high ($A_y=61\%$ at 5.05~GHz), limiting the cross-polarized power coefficient to $R_\mathrm{cross}=R_{xy}=37\%$ [Fig.~\ref{fig:ResultsPolarization}]. The remaining power is found in transmission (not shown), which is in all cases small due to the presence of the (perforated) copper backplane. To further quantify the performance we define the metrics of extinction ($\mathrm{Ext}=R_\mathrm{cross}/R_\mathrm{co}$) and efficiency ($\mathrm{Eff}=R_\mathrm{cross}/(1-A)$). They describe the completeness of the conversion in reflection and the percentage of cross-polarized reflection out the power that does not get absorbed, respectively. For the case of Fig.~\ref{fig:ResultsPolarization} their peaks appear at $\sim5.05$~GHz and equal $\mathrm{Ext}_{xy}=27$~dB and $\mathrm{Eff}_{xy}=97\%$, respectively.

Increasing the capacitance $C_1$ inside the available range, 2.6 to 3.5~pF, while keeping $R_1$ constant, we can shift the peak of cross-polarized reflection to lower frequencies. Specifically, the frequency variation can cover a range of 120~MHz. Along this frequency blueshift, we get a decrease in the cross-polarized reflected power, since the reactances of the two branches become more comparable; however, the efficiency and extinction remain in all cases very high.

Finally, extending the range of available $R$ and $C$ values by the chip to cover lower resistances, reducing the absorption, as well as a wider span of capacitances, to strengthen the reactance dissimilarity between the on- and off- diagonal elements, would further enhance the polarization conversion performance. The extension of $RC$ range will also enhance the reflected power of anomalous reflection. Alternatively, employing different unit cell topologies, the constraints on the IC range could be relaxed so as to be in line with the practical $RC$ values available by affordable chip designs.

\section{Conclusions and future prospects} 
\label{sec:4:Future}
Based on the introduced design procedure and the implementation constraints, we presented a fabrication-ready tunable microwave metasurface which offers quasi-continuous local control over the complex surface impedance and allows multiple reconfigurable functionalities. The dynamic control has been achieved by assembling a customized integrated-circuit chip in each unit cell and co-designing the electromagnetic and electronic components to fulfill the desired functionalities. \tblu{The structure's performance, primarily as a tunable perfect absorber, but also as a tunable anomalous reflector and a polarization converter, has been demonstrated, highlighting the multi-functional character of the proposed reconfigurable metasurface structure.} \tblu{Measured performance of this device will be published once the ICs have been fabricated, measured, and assembled on the manufactured boards.}

HyperSurfaces (HSF), combining the IC-enabled multi-functional reconfigurable metasurface aspects detailed in this work with appropriate hardware and software, enable themselves to act as hypervisors for metasurface functionalities. In other words, using well-defined application programming interfaces and communication protocols~\cite{LiaskosAPI,LiaskosComp}, a computer, a software service, or an individual can deploy and chain together multiple EM functionalities over the same HSF. For instance, complex alterations over the phase, amplitude and polarization can be deployed over a HSF at the same time. Recent studies have also modeled the wavefront detection as a software service, \tblu{allowing future HSFs to concurrently sense and modify} an impinging wave and enabling real-time adaptive operation~\cite{ABsense,CS.DCOSS.2019}.

The inter-networked deployment of multiple HSF units within a space yields a programmable wireless environment (PWE) as a whole, within which EM propagation becomes software-defined~\cite{metasurfacesNetworking}. For instance, applying HSF-coatings over walls, ceilings and other large surfaces in a floorplan, yields an indoors PWE. A central server can orchestrate the EM functionalities deployed to each HSF unit, matching it to the needs of mobile users present in the floorplan. In this manner, studies have showcased novel potential in PWE-assisted channel equalization, multipath fading, Doppler effect mitigation, long-range wireless power transfer and physical-layer security~\cite{phySecAdHoc}. These capabilities can enable novel, holistic data networking approaches, which will adaptively and jointly tune the data control logic and the physical propagation characteristics, to optimally serve the user needs. 


\section*{Acknowledgments}
The first three authors, A. P., O. T., and F. L., contributed equally to this work. 
{University of Cyprus researchers Loukas Petrou, Giorgos Varnava, and Petros Karousios are acknowledged for fruitful discussions.}
Odysseas Tsilipakos acknowledges the financial support of the Stavros Niarchos Foundation within the framework of the project ARCHERS (‘‘Advancing Young Researchers Human Capital in Cutting Edge Technologies in the Preservation of Cultural Heritage and the Tackling of Societal Challenges’’).

\bibliographystyle{IEEEtran}
\bibliography{IEEEabrv,refs_list}

\begin{IEEEbiography}[{\includegraphics[width=1in,height=1.25in,clip,keepaspectratio]{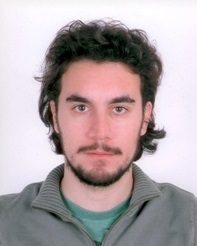}}]{Alexandros Pitilakis} received his diploma in Electrical Engineering from the Dept. of Electrical and Computer Engineering, Aristotle University of Thessaloniki (AUTH), Greece, in 2005, and his PhD from the same department in 2013. He holds an MSc degree in Electrical Engineering from the ENST (Telecom) Paris, 2007, including an internship at the Alcatel-Lucent Optical Transmission Systems group in Marcoussis, France. He is currently a postdoctoral researcher in AUTH, affiliated with the Foundation for Research and Technology Hellas (FORTH), and teaches undergraduate optics, photonics and antennas \& propagation courses in the University of Western Macedonia. His research interests include computational electromagnetics, waveguides and antennas (optical, THz, RF), metamaterials, nonlinear optics, integrated photonics, plasmonics, graphene. He is a member of OSA.
\end{IEEEbiography}

\begin{IEEEbiography}[{\includegraphics[width=1in,height=1.25in,clip,keepaspectratio]{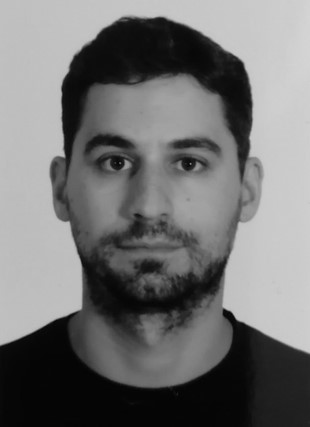}}]{Odysseas Tsilipakos} (S’06-M’14-SM’19) obtained the diploma and Ph.D. degrees from the Dept. of Electrical and Computer Engineering, Aristotle University of Thessaloniki (AUTh) in 2008 and 2013, respectively. From 2014 to 2015 he was a Postdoctoral Research Fellow with AUTh. Since 2016 he is a Postdoctoral Researcher at the Institute of Electronic Structure and Laser (IESL) in Foundation for Research and Technology Hellas (FORTH). His research interests span metasurfaces and metamaterials, plasmonics and nanophotonics, nonlinear optics in resonant and waveguiding structures, graphene and 2D photonic materials, and theoretical and computational electromagnetics. He is a Senior Member of IEEE and a member of The Optical Society (OSA).
\end{IEEEbiography}

\begin{IEEEbiography}[{\includegraphics[width=1in,height=1.25in,clip,keepaspectratio]{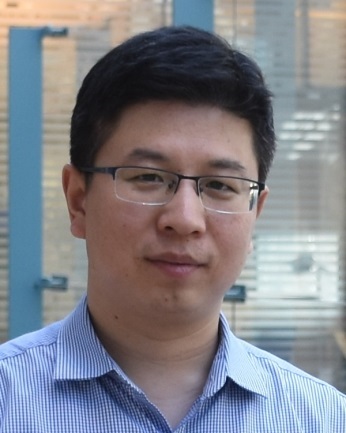}}]{Fu Liu} received the B.Sc. degree in applied physics from the China University of Mining and Technology, Xuzhou, China, in 2008, the M.Sc. degree in theoretical physics from Beijing Normal University, Beijing, China, in 2011, and the Ph.D. degree in physics from the City University of Hong Kong, Hong Kong, China in 2015. From 2014 to 2016, he was a Research Fellow with the School of Physics and Astronomy, University of Birmingham, Birmingham, U.K. Since 2017, he has been a Post-Doctoral Researcher with the Department of Electronics and Nanoengineering, School of Electrical Engineering, Aalto University, Espoo, Finland.
\end{IEEEbiography}

\begin{IEEEbiography}[{\includegraphics[width=1in,height=1.25in,clip,keepaspectratio]{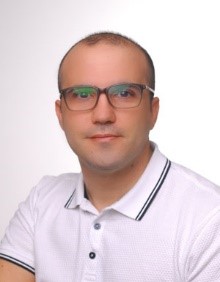}}]{Kypros M. Kossifos} (S’16-M’20) was born in Larnaca, Cyprus. He received the diploma in electronic engineering from the ATEI of Athens, Athens, Greece in 2010 and received the MSc in electrical engineering from the University of Cyprus, Nicosia, Cyprus in 2015. He is currently a PhD Candidate at the University of Cyprus. In 2017 he began working on the FETOPEN project VISORSURF (A Hardware Platform for Software-driven Functional Metasurfaces). His research interests include metamaterials, RF/microwave circuits and antennas. 
\end{IEEEbiography}

\begin{IEEEbiography}[{\includegraphics[width=1in,height=1.25in,clip,keepaspectratio]{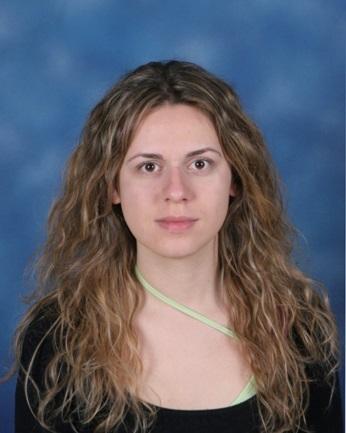}}]{Anna C. Tasolamprou} received her Diploma and PhD degrees in Electrical and Computer Engineering from the Aristotle University of Thessaloniki (AUTH), Greece. Her interests lie in the area of electromagnetics with a focus on the wave propagation properties through random and periodic media such as photonic crystals, liquid crystals, 2D materials, metamaterials and metasurfaces. She has a long experience in analytical and numerical methods for electrodynamics and an active presence in contemporary applied physics. She is currently working in the Photonic-Phononic and Meta-Materials (PPM) Group of the Institute of Electronic structure and Laser (IESL) of the Foundation for Research and Technology Hellas (FORTH).
\end{IEEEbiography}

\begin{IEEEbiography}[{\includegraphics[width=1in,height=1.25in,clip,keepaspectratio]{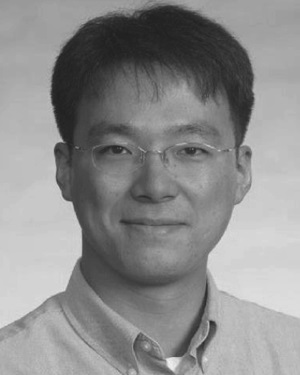}}]{Do-Hoon Kwon} (S’94–M’00–SM’08) received the B.S. degree in electrical engineering from Korea Advanced Institute of Science and Technology, Daejeon, South Korea, in 1994, and the M.S. and Ph.D. degrees in electrical engineering from Ohio State University, Columbus, OH, USA, in 1995 and 2000, respectively. He was a Senior Engineer with Samsung Electronics, South Korea, from 2000 to 2006. From 2006 to 2008, he was a Post-Doctoral Researcher with the Material Research Science and Engineering Center and the Department of Electrical Engineering, Pennsylvania State University, University Park, PA, USA. In 2008, he joined the Department of Electrical and Computer Engineering, University of Massachusetts Amherst, Amherst, MA, USA. He was a Summer Faculty Fellow with the Sensors Directorate, Air Force Research Laboratory, Wright-Patterson AFB, OH, USA, in 2011. His current research interests include antenna scattering theory, small/wideband antennas and array elements, frequency selective surfaces, metamaterials, cloaking, and transformation electromagnetic/optical device designs. Dr. Kwon was a recipient of the inaugural IEEE Antennas and Propagation Society Edward E. Altshuler Prize Paper Award in 2011.
\end{IEEEbiography}

\begin{IEEEbiography}[{\includegraphics[width=1in,height=1.25in,clip,keepaspectratio]{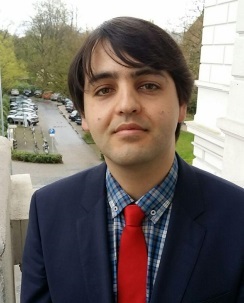}}]{Mohammad Sajjad Mirmoosa} was born in Baft, Iran. He received the Doctor of Science (Technology) degree from Aalto University in August 2017. Currently, he has a postdoctoral researcher position, and his main research interests are electromagnetic theory, light-matter interaction and radiative heat transfer.
\end{IEEEbiography}

\begin{IEEEbiography}[{\includegraphics[width=1in,height=1.25in,clip,keepaspectratio]{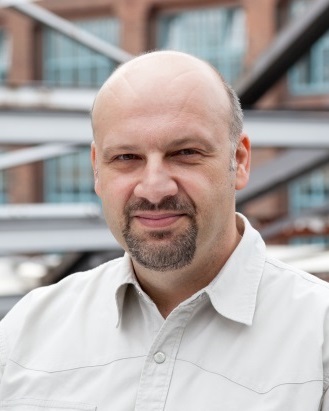}}]{Dionysios Manessis} possesses M.Sc. and Ph.D degrees in Materials Science \& Engineering from Stevens Institute of Technology, NJ, USA and project leadership certificate degrees from Cornell University, NY, USA. He has worked as Technologist for Universal Instruments Corporation in NY, USA and since 2001 has been Senior Technology Scientist in Fraunhofer IZM in Berlin. His main research interests lie on fine-pitch flip-chip and wafer-level CSP bumping, solder balling, materials selection for advanced packaging technologies, embedding processes for heterogeneous integration of components in PCBs and optical PCBs, large scale prototype manufacturing. In the above technical fields, he has published extensively in international conferences and peer-reviewed journals. 
\end{IEEEbiography}

\begin{IEEEbiography}[{\includegraphics[width=1in,height=1.25in,clip,keepaspectratio]{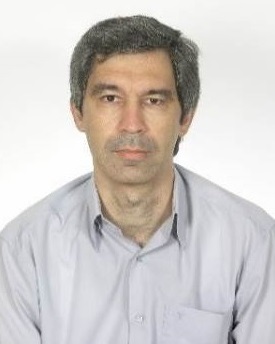}}]{Nikolaos V. Kantartzis} is a professor of computational electromagnetics and electromagnetic compatibility in the School of Electrical and Computer Engineering at the Aristotle University of Thessaloniki (AUTH), Greece, from where he received the Diploma and Ph.D. degrees in 1994 and 1999, respectively. He has authored/coauthored 3 books, 9 book chapters, more than 135 peer-reviewed journal papers and 230 conference papers. His primary research interests include computational electromagnetics, metamaterials, graphene, real-world EMC/EMI problems, microwaves and antennas as well as nanotechnology devices. Dr. Kantartzis is a Senior Member of IEEE, an ICS and ACES member.
\end{IEEEbiography}

\begin{IEEEbiography}[{\includegraphics[width=1in,height=1.25in,clip,keepaspectratio]{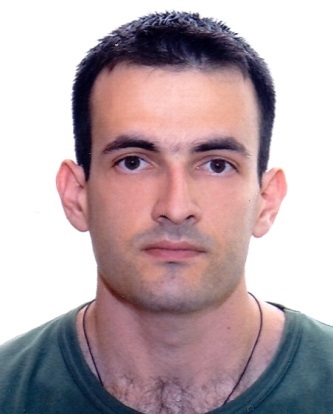}}]{Christos Liaskos} received the Diploma in Electrical Engineering from the Aristotle University of Thessaloniki (AUTH), Greece in 2004, the MSc degree in Medical Informatics in 2008 from the Medical School, AUTH and the PhD degree in Computer Networking from the Dept. of Informatics, AUTH in 2014. He is currently a post-doc researcher at the Foundation of Research and Technology, Hellas (FORTH), handling the scientific and technical co-coordination task of project VISORSURF (www.visorsurf.eu). His research interests include computer networks, traffic engineering and novel control schemes for wireless communications.
\end{IEEEbiography}

\begin{IEEEbiography}[{\includegraphics[width=1in,height=1.25in,clip,keepaspectratio]{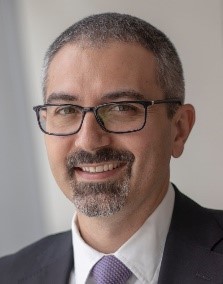}}]{Marco A. Antondiades} (S’99–M’09–SM’17) received the B.A.Sc. degree in electrical engineering from the University of Waterloo, ON, Canada, in 2001, and the M.A.Sc. and Ph.D. degrees in electrical engineering from the University of Toronto, ON, Canada, in 2003 and 2009, respectively. His research interests include engineered electromagnetic materials (metamaterials, metasurfaces), electrically-small antennas, adaptive and reconfigurable antennas, RF/microwave circuits and devices, implantable/wearable antennas, microwave imaging, wireless power transfer, and radio-frequency identification. Dr. Antoniades was a recipient of the Hellenic Canadian Federation of Ontario Academic Excellence Award in 2003, the First Prize in the Student Paper Competition at the 2006 IEEE AP-S International Symposium on Antennas and Propagation (AP-S/URSI), a Best of IEEE Computer Architecture Letters Award in 2018, and a Teaching Innovation Award from the University of Cyprus in 2018. He served as an Associate Editor of IET Microwaves, Antennas and Propagation from 2014 to 2018, as the Conference Chair of the 2015 Loughborough Antennas and Propagation Conference (LAPC), and on the Steering Committee for the 2010 IEEE AP-S/URSI International Symposium. He is a member of the IEEE AP-S Education Committee, and serves as a Co-Chair of the Technical Program Committee for the 2020 AP-S/URSI International Symposium. He is an Associate Editor of the IEEE Transactions on Antennas and Propagation, and the IEEE Antennas and Wireless Propagation Letters.
\end{IEEEbiography}

\begin{IEEEbiography}[{\includegraphics[width=1in,height=1.25in,clip,keepaspectratio]{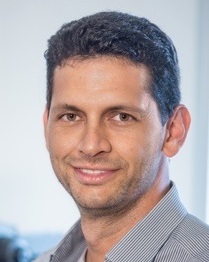}}]{Julius Georgiou} (M’98-SM’08) is an Associate Professor at the University of Cyprus. He received his M.Eng degree in Electrical and Electronic Engineering and Ph.D. degree from Imperial College London in 1998 and 2003 respectively. For two years he worked as Head of Micropower Design in a technology start-up company, Toumaz Technology. In 2004 he joined the Johns Hopkins University as a Postdoctoral Fellow, before becoming a faculty member at the University of Cyprus from 2005 onwards.
Prof. Georgiou is a member of the IEEE Circuits and Systems Society, was the Chair of the IEEE Biomedical and Life Science Circuits and Systems (BioCAS) Technical Committee, as well as a member of the IEEE Circuits and Systems Society Analog Signal Processing Technical Committee. He served as the General Chair of the 2010 IEEE Biomedical Circuits and Systems Conference and was the Action Chair of the EU COST Action ICT-1401 on “Memristors-Devices, Models, Circuits, Systems and Applications - MemoCIS”. Prof. Georgiou was an IEEE Circuits and Systems Society Distinguished Lecturer for 2016-2017. He is also is an Associate Editor of the IEEE Transactions on Biomedical Circuits and Systems and Associate Editor of the Frontiers in Neuromorphic Engineering Journal. He is a recipient of a best paper award at the IEEE ISCAS 2011 International Symposium and at the IEEE BioDevices 2008 Conference. In 2016 he received the 2015 ONE Award from the President of the Republic of Cyprus for his research accomplishments.
\end{IEEEbiography}

\begin{IEEEbiography}[{\includegraphics[width=1in,height=1.25in,clip,keepaspectratio]{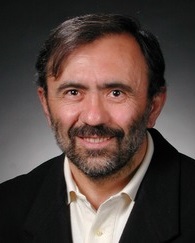}}]{Costas M. Soukoulis} obtained his Ph.D. in 1978 at the Physics Department of University of Chicago. He is currently a senior scientist at FORTH and Distinguished Professor of Physics at Iowa State University. He is a Fellow of the APS, OSA and AAAS. He has more than 400 publications in refereed journals and he has given more than 173 invited talks. His work on photonic band gaps (PBG), random lasers, plasmonics, graphene, metasurfaces and nonlinear systems is well known and well received. He has organized three NATO ASI on PBGs and he was the director of PECS-VI that took place in Crete in June 2005. He also has five patents concerning the potential applications of the photonic band gaps and left-handed materials. He recently received the Senior Humboldt Research Award. He has been cited ~57000 times in SCI journals. He is the winner of the 2014 Max Born Award given by the Optical Society of America "for his creative and outstanding theoretical and experimental research in the fields of photonic crystals and left-handed metamaterials." He was included in the 2014 and 2015 Highly Cited Researchers by Thompson-Reuters. In 2013 he won the McGroddy Prize for New Materials “For the discovery of metamaterials”, awarded by the American Physical Society. He has lead R\&D projects funded by the EU, NSF, DOE, DARPA, the Office of Naval Research, MUTI-AFOSR, NATO, EPRI and Ames Lab  
\end{IEEEbiography}

\begin{IEEEbiography}[{\includegraphics[width=1in,height=1.25in,clip,keepaspectratio]{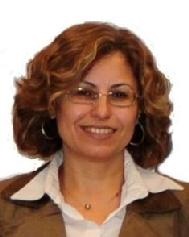}}]{Maria Kafesaki} is Associate Professor in the Dept. of Materials Science and Technology of the University of Crete and associated Researcher at the Institute of Electronic Structure and Laser (IESL) of Foundation for Research and Technology Hellas (FORTH). She obtained her Ph.D. in 1997, at the Physics Department of the University of Crete, Greece. She has worked as post-doctoral researcher in the Consejo Superior de Investigaciones Cientificas in Madrid, Spain, and in IESL of FORTH. (1997-2001). Her current research is on the area of electromagnetic wave propagation in periodic and random media, with emphasis on photonic crystals and metamaterials, especially negative refractive index materials. She has around 130 publications in refereed journals and conference proceedings (with ~9200 citations), she has participated in various European projects as well as in the organization of many international conferences and schools. Also she is Fellow of the Optical Society of America
\end{IEEEbiography}

\begin{IEEEbiography}[{\includegraphics[width=1in,height=1.25in,clip,keepaspectratio]{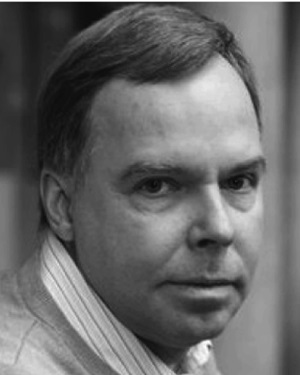}}]{Sergei A. Tretyakov} (M’92–SM’98–F’08) received the Dipl.Ing.-Physicist, Ph.D., and D.Sc. degrees in radiophysics from Saint Petersburg State Technical University, Saint Petersburg, Russia, in 1980, 1987, and 1995, respectively. From 1980 to 2000, he was with the Radiophysics Department, Saint Petersburg State Technical University. He is currently a Professor of radio science with the Department of Electronics and Nanoengineering, Aalto University, Espoo, Finland. He has authored or co-authored five research monographs and over 270 journal papers. His current research interests include electromagnetic field theory, complex media electromagnetics, metamaterials, and microwave engineering. Dr. Tretyakov served as the General Chair for the International Congress Series on Advanced Electromagnetic Materials in Microwaves and Optics (Metamaterials) from 2007 to 2013, the Chairman for the Saint Petersburg IEEE Electron Devices/Microwave Theory and the Techniques/Antennas and Propagation Chapter from 1995 to 1998, and the President for the Virtual Institute for Artificial Electromagnetic Materials and Metamaterials (Metamorphose VI) from 2007 to 2013.
\end{IEEEbiography}

\end{document}